\documentclass[onecolumn,A4]{article}
\usepackage{cite}
\usepackage{anysize}
\usepackage{mathrsfs}
\usepackage{amsmath}
\usepackage{amssymb}
\usepackage{graphicx}
\usepackage{subfigure}
\usepackage{amssymb}
\usepackage{amsthm}
\pdfoutput=1
\usepackage{pdfpages}
\DeclareGraphicsExtensions{.jpeg}

\begin{document}

\vskip 1cm
\marginsize{3cm}{3cm}{3cm}{1cm}

\begin{center}
{\bf{\huge Experimental and numerical simulation of a TPC like set up for the measurement of ion backflow}}\\
~\\
Deb Sankar Bhattacharya$^{a,b,c}$, Purba Bhattacharya$^d$, Prasant Kumar Rout$^a$, Supratik Mukhopadhyay$^{a*}$, Sudeb Bhattacharya$^e$, Nayana Majumdar$^a$, Sandip Sarkar$^a$, Paul Colas$^b$, David Attie$^b$, Serguei Ganjour$^b$, Aparajita Bhattacharya$^c$\\
{\em $^a$ Applied Nuclear Physics Division, Saha Institute of Nuclear Physics, Kolkata - 700064, India}\\
{\em $^b$ IRFU, CEA, Université Paris-Saclay, F-91191 Gif sur Yvette, France}\\
{\em $^c$ Department of Physics, Jadavpur University, Jadavpur, Kolkata - 700032, India}\\
{\em $^d$ Department of Particle Physics and AstroPhysics, Weizmann Institute of Science, Herzl St. 234, Rehovot - 7610001, Israel}\\
{\em $^e$ Retired Senior Professor, Applied Nuclear Physics Division, Saha Institute of Nuclear Physics, Kolkata - 700064, India}\\
~\\
~\\
~\\
~\\
~\\
{\bf{\large Abstract}}
\end{center}

\noindent Ion backflow is one of the effects limiting the operation of a 
gaseous detector at high flux, by giving rise to space charge which 
perturbs the electric field.
The natural ability of bulk Micromegas to suppress ion feedback
is very effective and can help the TPC drift volume to remain 
relatively free of space charge build-up.
An efficient and precise measurement of the backflow fraction is 
necessary to cope up with the track distortion due to the space 
charge effect.
In a subtle but significant modification of the usual approach, 
we have made use of two drift meshes in order to measure the ion 
backflow fraction for bulk Micromegas detector. This helps to truly represent the backflow fraction for a TPC. Moreover, attempt is taken to optimize the field configuration between the drift meshes.
In conjunction with the experimental measurement, Garfield simulation
framework has been used to simulate the related physics processes numerically.

\vskip 1.5cm
\begin{flushleft}

\end{flushleft}

\vskip 1.5in
\noindent
{\bf ~$^*$Corresponding Author}: Supratik Mukhopadhyay

E-mail: supratik.mukhopadhyay@saha.ac.in

\newpage

\section{Introduction}

\begin{figure}[hbt]
\centering
\includegraphics[scale=0.25]{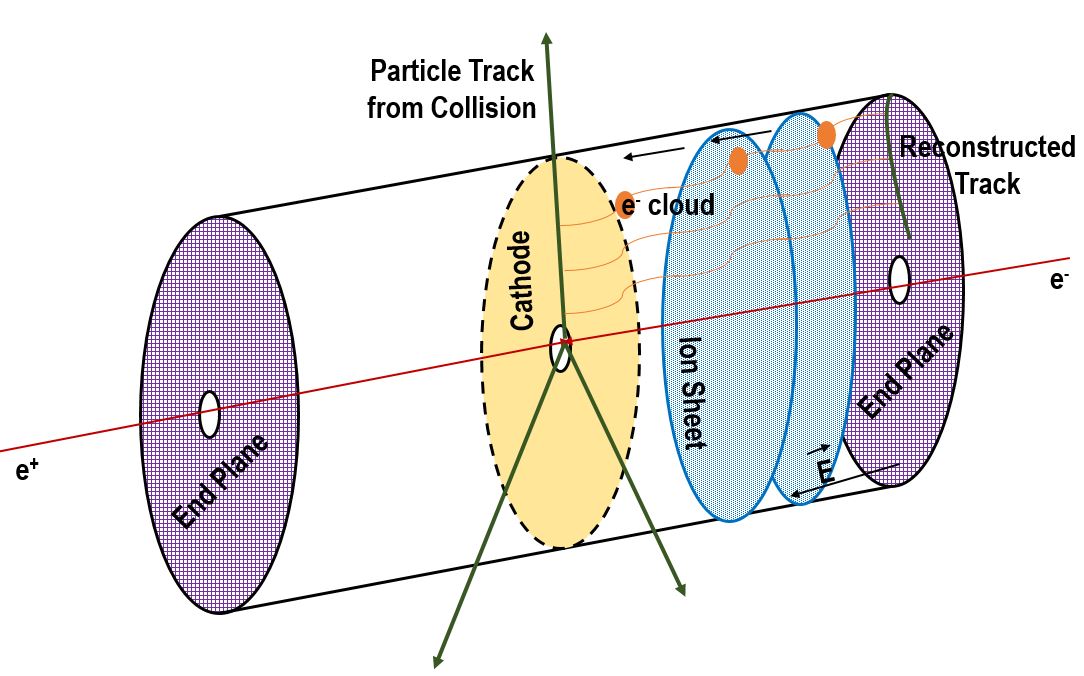}
\caption{Schematic diagram of the TPC operated in the collider experiment. 
Ion discs in the TPC under normal operation is shown here.}
\label{TPC}
\end{figure}

The International Linear Collider (ILC) \cite{ILC} is a proposed electron-positron collider for Higgs precision measurements and discovery.
It aims for the physics studies complementary to the Large Hadron Collider (LHC).
The goals of physics studies at the ILC have pushed the requirements for the 
detector to an unprecedented level.
These requirements include good momentum resolution, high jet energy 
resolution and excellent particle identification.
The International Large Detector (ILD) \cite{ILD} is one of the two concepts for the ILC.
This detector concept is optimized for the particle flow reconstruction which 
requires a highly efficient tracking system.
The central tracker of the ILD is conceived to be a Time Projection Chamber (TPC) \cite{TPC} 
(Fig.~\ref{TPC}) which will reconstruct the three-dimensional tracks of the charged particles.
A TPC has the advantage of small material budget, truly continuous tracking and robust pattern recognition.

The classical readout system in TPCs are the Multi Wire Proportional Chambers 
(MWPC). 
However, due to the inherent restrictions associated with the design / 
construction of MWPC, the wire planes can not fulfill the ILD
requirements.
For example, a large-volume TPC based on wire readout will suffer from the 
space charge effect originating from the high fluxes of the backflowing ions.
In addition, the strong magnetic field of $\mathrm{B = 3.5T}$ and the wide gap of
$1-2~\mathrm{mm}$ between wires will lead to strong
$\vec{\mathrm{E}}\times\vec{\mathrm{B}}$
effects.

\begin{figure}[hbt]
\centering
\includegraphics[scale=0.25]{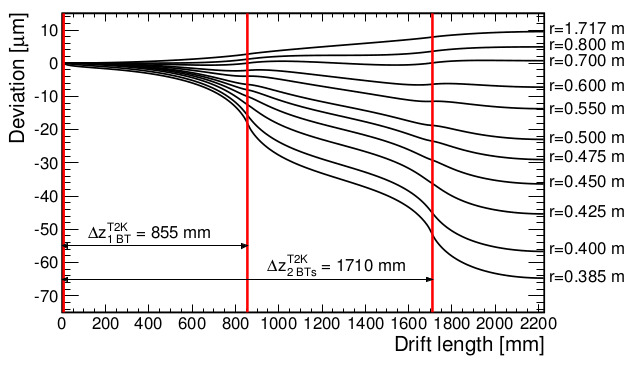}
\caption{Displacement of electron track due to positive ions \cite{Report}.}
\label{Distortion}
\end{figure}

As far as the ions are concerned, they can be divided into two types, the primary and the secondary ions.
The primary ions come from the ionization by the charged particle and move to
the cathode, whereas the secondary ions come from the gas amplification and migrate to the drift volume after getting partially collected.
In an experiment where the event rate is high, it is very important to limit the ion backflow from
the amplification region.
If not, the backflowing ions will cause significant non-uniformity of the electric field in drift volume leading to a distortion of the charged particle tracks.
According to certain estimates, the number of these secondary
ions is about 2 to 9 times larger than that of the primary ions and the distribution is different from the primary ones \cite{AsianGroup}.
Since the gain is expected to be around 5000, it is likely that the authors assumed significant (99.9\% if we consider 5 ions out of 5000) partial backflow suppression. 
The distortion of track by positive ions, according to this reference, is $60~\mu\mathrm{m}$ due to primary ions and $60~\mu\mathrm{m}$ due to the secondary ions.
The conclusion of this paper is that the necessary ion feedback should be smaller than $10^{-3}$ and the ion gate should remain open for 1ms and close following 199ms.
On the other hand, simulations carried out in \cite{Report, ThorstenPhDThesis} indicates an even more bleak future.
These calculation also assume that due to the bunch-train structure of the beam of ILC (one 1 ms train in every 
200 ms), the ions from the amplification zone will be concentrated in the form of discs of 
about 3 mm thickness near the readout (in the earlier report, this disk was assumed to be of 4mm thickness), and then flow back into the drift 
volume. 
There would be two / three such discs in the chamber during normal operation 
(Fig.\ref{TPC}).
The presence of such ion clouds would effect in a track distortion as large as 
$60~\mu\mathrm{m}$ (Fig.~\ref{Distortion}) for $e^+e^-$ pairs even if we consider 100\% suppression of secondary ion backflow, i.e., only one back drifting ion for every drift electron.
The use of an active gating grid (GG) for the ILD-TPC has not yet been finalized.
However, despite the disadvantage of introducing a dead time to the TPC, it can be seen as a feasible solution as reported in \cite{AsianGroup, Report, ThorstenPhDThesis}.
Moreover, corrections at the reconstruction stage is also an option that is being actively evaluated by several groups.
In any case, it is self-evident from the above discussion that ion backflow is a crucial issue for the ILD-TPC and is worthy of detailed investigation.

Micro-Pattern Gaseous Detectors (MPGDs) \cite{MPGD} provide good  
ion feedback suppression due to highly asymmetric electric fields 
in-between the drift and amplification regions. 
As a result, they are suitable for high luminosity experiments. 
Due to their very good position and time resolution, the MPGDs are expected
to meet the ambitious demands for the ILD. 
The $\mathrm{R}\&\mathrm{D}$ activities for the ILD TPC are currently concentrated on the adoption of the 
micro-pattern devices for the gas amplification stage.
Among different MPGDs, the Micro Mesh Gaseous Structure (Micromegas) 
\cite{Micromegas} fulfills the needs of high-luminosity 
colliders with increased reliability in harsh radiation environments.

\begin{figure}[hbt]
\centering
\includegraphics[scale=0.4]{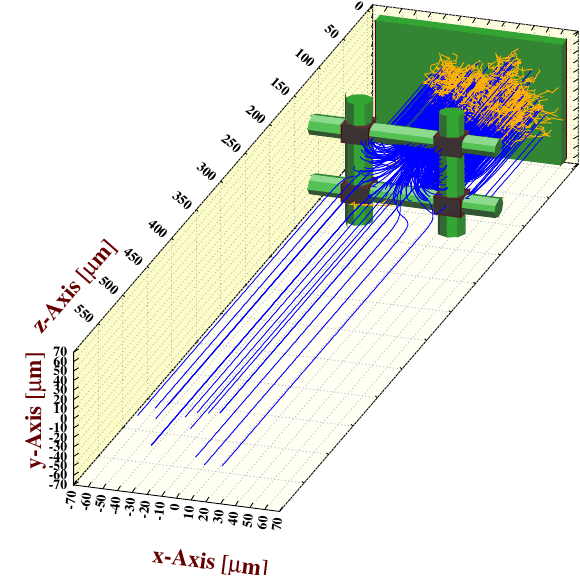}
\caption{Electron avalanche and ion drift lines of a Micromegas detector.}
\label{IBF-1}
\end{figure}

The Micromegas is a parallel plate device 
and composed of a very thin metallic micro-mesh, which separates the 
low-field drift region from the high-field amplification region.
Due to the field gradient between the drift and amplification regions and 
the periodic hole pattern, the field lines from the drift region are 
compressed in the vicinity of the micro-mesh holes and form funnels 
having width of a few microns in the amplification region \cite{Paul}.
As a result, an electron approaching the micro-mesh is focused towards 
the center of a hole and produces an avalanche inside the funnel.
Due to the transverse diffusion, the avalanche also extends outside the 
funnel.
Conversely, the ions, due to their larger mass, are not affected much by 
the diffusion and drift along the field lines.
Most of the ions, created outside the funnel, follow the field lines 
and are collected by the micro-mesh (Fig.\ref{IBF-1}).
A very small fraction, produced in the thin funnel, drifts back towards 
the drift volume.
The ion backflow fraction can, thus, be defined as:
\begin{eqnarray}
IBF = \frac {N_b} {N_t}
\end{eqnarray}
\noindent where $N_t$  is the average number of ions produced in an 
electron avalanche and $N_b$ the average number of the backflowing ions.
For Micromegas, this fraction is very small, but can still be significant 
with a high track density.
Therefore, a proper estimation of the backflow fraction is necessary 
to carry out possible design modifications and to develop algorithms for 
the possible correction of track distortion.

We have tried to experimentally simulate the ion back flow in a TPC 
using a small setup. 
For this purpose, experimental studies have been carried out to measure the 
ion backflow fraction for bulk Micromegas detectors \cite{Bulk} in two Argon based
gas mixtures, namely, $\mathrm{Argon}$+$\mathrm{Isobutane}$ ($95:5$) and
$\mathrm{Argon}$+$\mathrm{CF_4}$+$\mathrm{Isobutane}$ ($95:3:2$).
In order to remain true to a TPC environment, we have used two drift 
meshes as described in \cite{Purba1, Purba2}.
In these references, we had also described the details of the measurement
and compared them with numerical estimates to understand the dependence of
the backflow fraction on the detector design parameters.
In this work, we dwell upon the optimization of the experimental setup
itself and also present the variation of IBF for two gas mixtures.
In conjunction with the experimental work, extensive numerical simulations 
related to the optimization of the experimental setup have been carried out
using Garfield \cite{Garfield1, Garfield2}.

\section{Experimental Setup}
\label{sec:experiment}

As already discussed in our earlier work \cite{Purba1, Purba2}, the use 
of a single drift mesh is likely to lead to erroneous estimates of ion backflow
occurring in a real TPC.
For example, in the ILD TPC, the cathode is planned to be placed in 
the middle, dividing the whole chamber into two equal halves (Fig.~\ref{TPC}) \cite{ILD}.
The primary ionization occurs in the gas volume that consists of the drift and the amplification regions. 

\begin{figure}
\centering
\subfigure[]
{\includegraphics[scale=0.4]{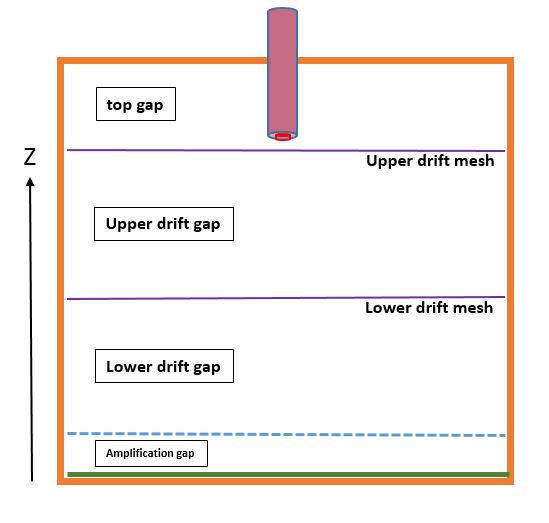}}
\subfigure[]
{\includegraphics[scale=0.3]{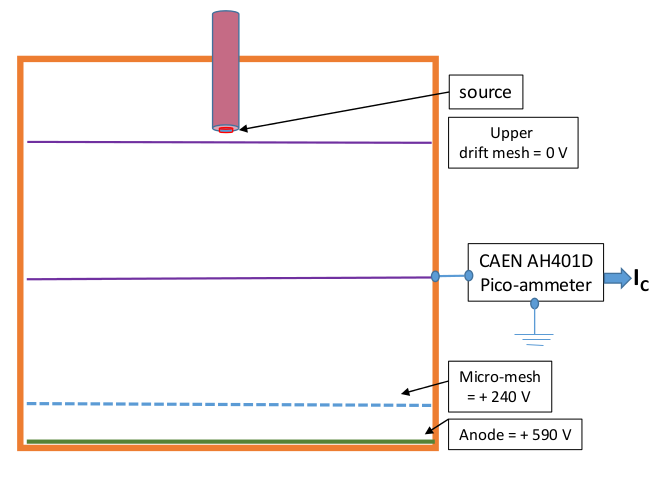}}
\caption{(a) Simple schematic diagram of experimental setup for ion backflow measurement, (b) Configuration for measuring current from the first drift mesh.}
\label{ExptSetup}
\end{figure}

\begin{figure}
\centering
\subfigure[]
{\label{Testbox-1}\includegraphics[scale=0.06]{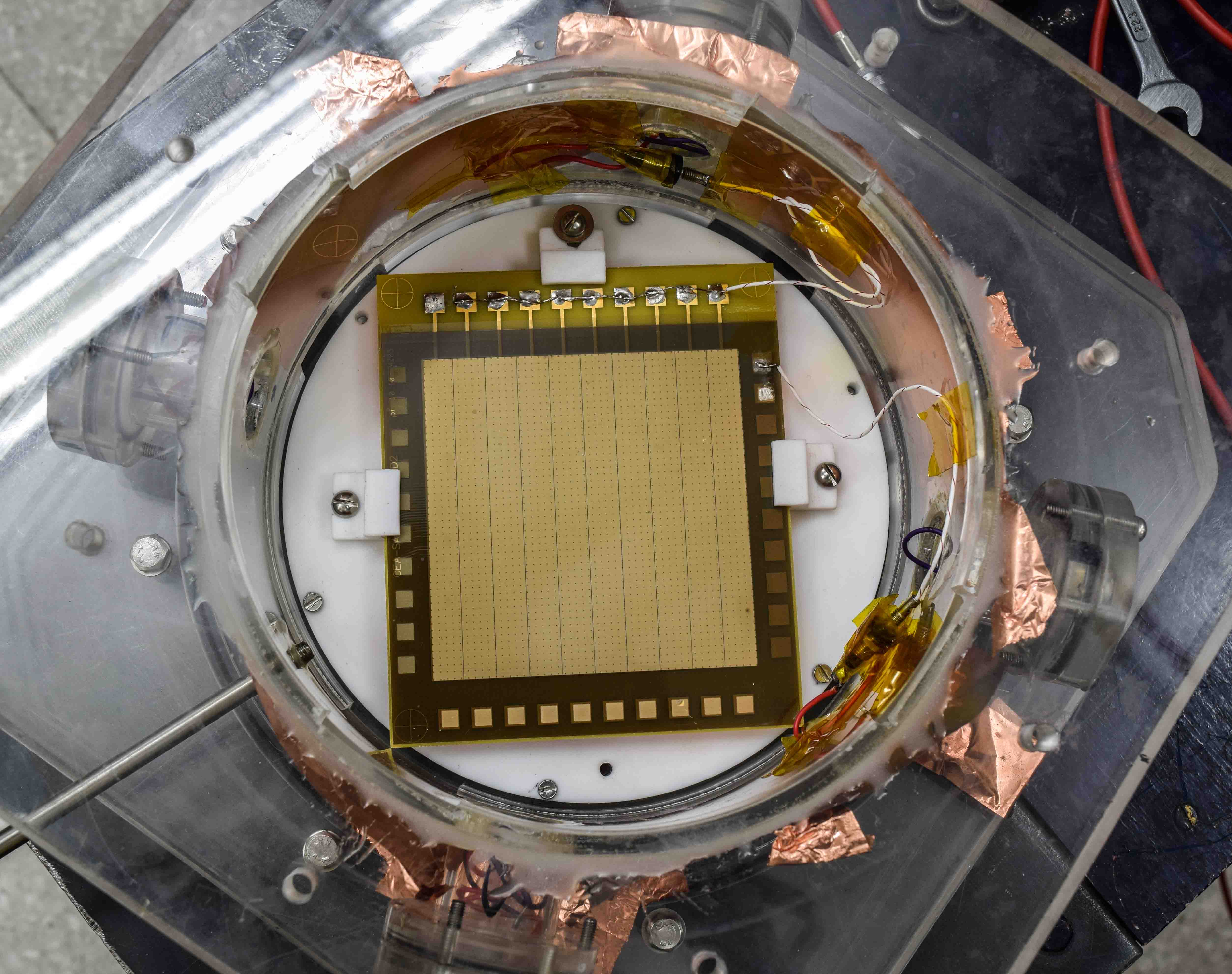}}
\subfigure[]
{\label{Testbox-2}\includegraphics[scale=0.07]{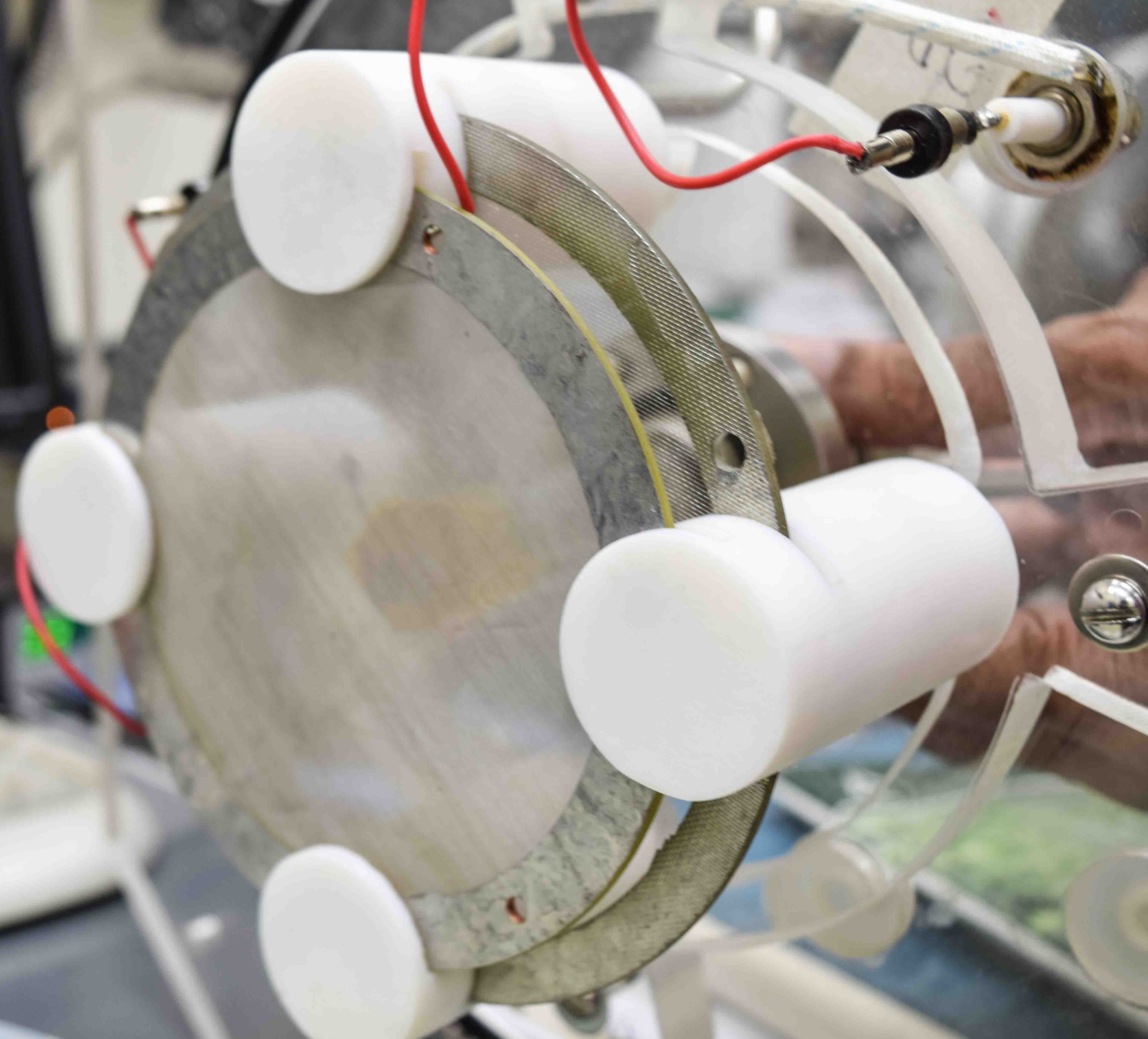}}
\caption{Photograph of (a) the test box with the detector and two drift mesh, (b) the whole setup.}
\label{Testbox}
\end{figure}

However, in the laboratory experimental setup, along with the contribution of 
the ions from the amplification and the drift region, there is a possibility of 
having an additional contribution to the drift current from the ions created 
in Regions 2 and 3 in Fig.\ref{ExptSetup} (the region in between the first drift mesh and 
the window of the test box).
These additional ions are non-existent in a real TPC since there is no 
such volume there (Fig.~\ref{TPC}). 
Thus, in the laboratory, correct estimation of IBF is difficult using a setup that has
only the $1^{st}$ drift mesh. 
So the setup has been modified by placing a second drift mesh
at a distance of $1~\mathrm{cm}$ above the first one.
The ions that are created between the test box window and the $2^{nd}$ drift 
mesh (Region 3 in Fig.\ref{ExptSetup}) are collected on the outer drift mesh ($2^{nd}$ 
Drift Mesh in Fig.\ref{ExptSetup}).
The ions in Region 2 will be collected by either the $1^{st}$ or the $2^{nd}$ drift
mesh, depending on the electric field configuration.
Thus, the voltages should be applied so as to ensure that the 1st drift mesh collects
ions only from the Drift Region, while the 2nd drift mesh collects ions from Regions 2
and 3.
Only in such an event, the current on the inner drift mesh ($1^{st}$ Drift Mesh in
Fig.\ref{ExptSetup}) will provide a reasonable estimate of the ionic current from the 
avalanche.
The photographs of the test box using two drift meshes and the positioning of 
the source are shown in Fig.\ref{Testbox}.

The currents on the drift mesh and the micro-mesh have been measured and the 
ratio of these two currents gives an estimate of the backflow fraction.
\begin{eqnarray}
IBF =  \frac {I_C} {(I_M + I_C)}
\end{eqnarray}
\noindent where $I_C$ is the current measured on the $1^{st}$ drift mesh 
(Fig.\ref{ExptSetup}) and is proportional to the number of ions collected on the drift 
mesh; $I_M$ is the current measured in the micro-mesh and proportional to the 
number of ions collected on the mesh.
For the measurement of current, a pico-ammeter (CAEN model AH401D) has been 
used (Fig.\ref{ExptSetup}) which can measure the current only from an electrode which is 
at a ground potential.
Because of this, the potential configuration, in the present experiment, 
has been suitably altered depending on the nature of measurement.
For example, in one configuration, for the measurement of current from the 
mesh, the micro-mesh has been grounded, whereas the anode plane and the 
drift plane have been biased with more positive and more negative voltages, 
respectively, with respect to the mesh plane, maintaining the proper field 
configuration.
Similarly, for measuring $I_C$, the drift plane is set to the ground 
potential and the mesh and the anode to the required positive high voltages.

\section{Simulation Conditions}

Garfield \cite{Garfield1, Garfield2} simulation framework has been used 
to understand the field configurations and its effects for various possible
relative voltage configuration of the two drift meshes.
The 3D electrostatic field simulation has been carried out using neBEM 
\cite{neBEM1, neBEM2, neBEM3} toolkit.
Besides neBEM, HEED \cite{HEED1, HEED2} has been used for primary 
ionization calculation and 
Magboltz \cite{Magboltz1, Magboltz2} for computing the transport proerties.

\begin{figure}[hbt]
\centering
\includegraphics[scale=0.3]{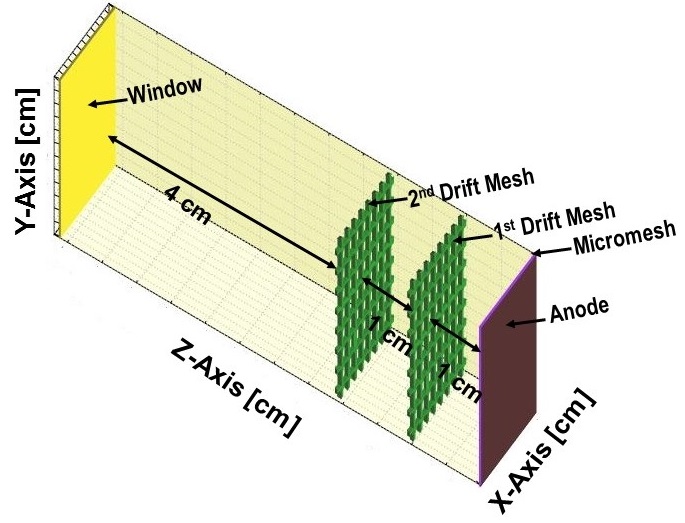}
\caption{Simulated model using two drift meshes.}
\label{SimulationModel}
\end{figure}

The micromesh and the two drift meshes have been modeled using Garfield 
as shown in Fig.~\ref{SimulationModel}.
A grounded anode plane $128~\mu\mathrm{m}$ below the micro-mesh and 
a dielectric window $4~\mathrm{cm}$ above the $2^{nd}$ drift mesh have
been considered. 
For this calculation, the primary ionization due to 5.9 keV photon track has 
been estimated using HEED. 
Then, these primary electrons have been made to drift towards the amplification 
region where they are allowed to get amplified.
The drift of the primary ions and the ions created in the avalanche have been 
traced.
The backflow fraction has been calculated as

\begin{eqnarray}
IBF = \frac {N_{id}} {(N_{id}+N_{im})}
\end{eqnarray}

\noindent where ${{N}_{id}}$ is the number of ions collected at the 
$1^{\mathrm{st}}$ drift mesh and ${{N}_{im}}$ is the number of ions 
collected at the micro-mesh.
It should be mentioned here that we have considered only thermal diffusion of ions while estimating IBF.

\section{Results}

\subsection{Numerical Discussion}

\begin{figure}[hbt]
\centering
\subfigure[]
{\label{Field-Single}\includegraphics[scale=0.05]{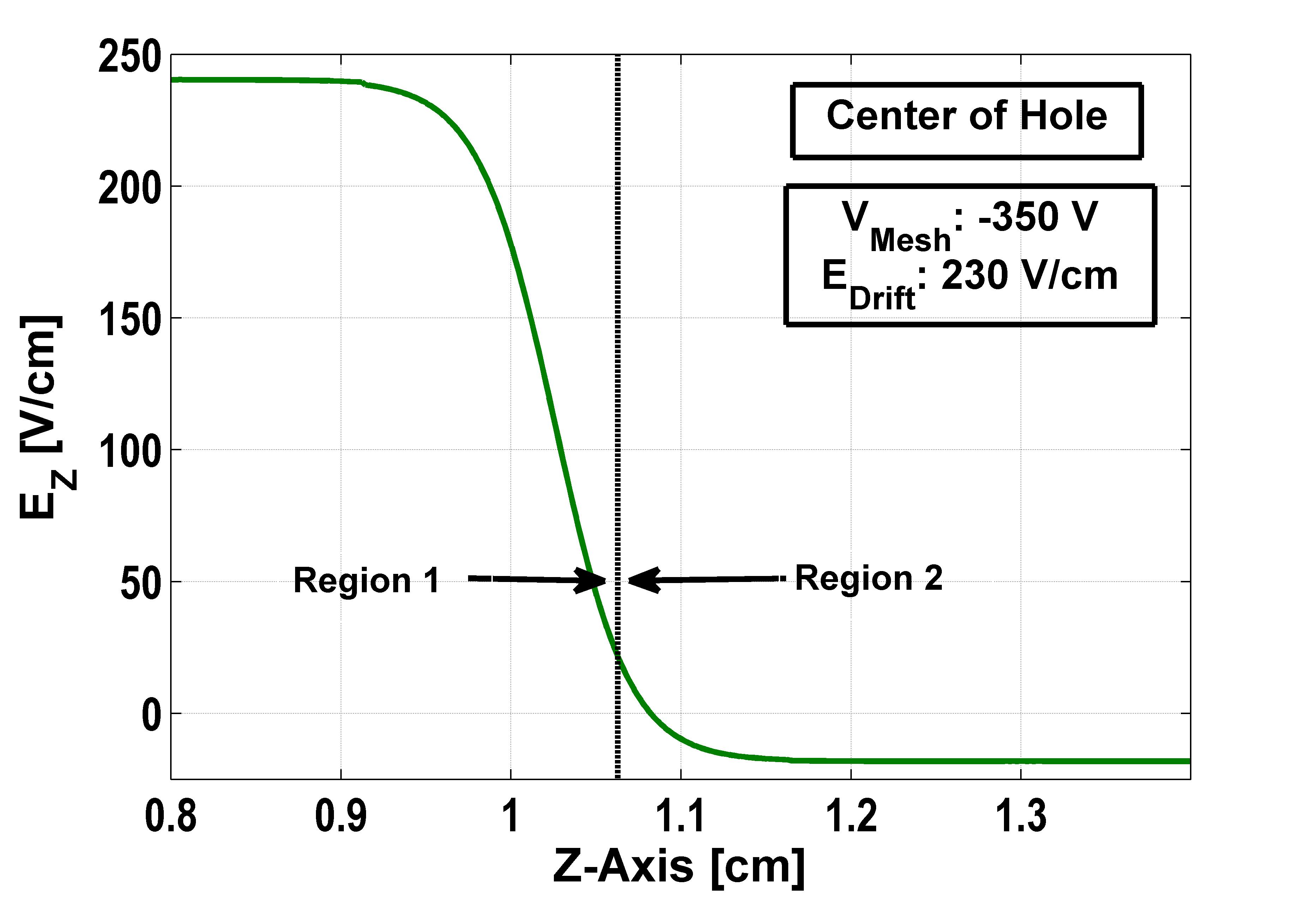}}
\subfigure[]
{\label{Drift-Single}\includegraphics[scale=0.35]{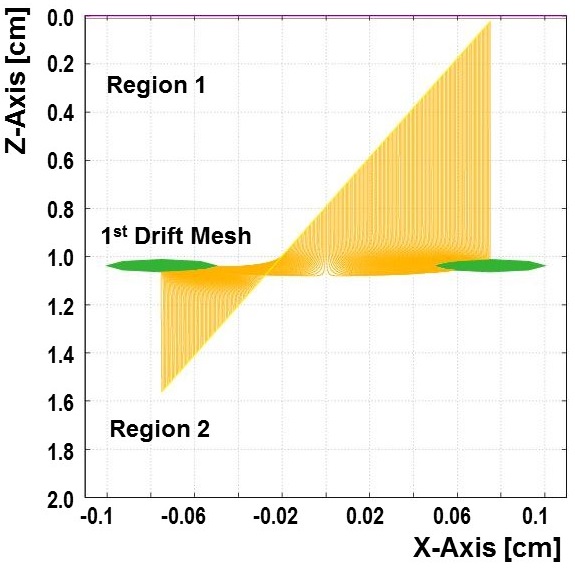}}
\caption{(a) Axial electric field in Region 1 and Region 2, (b) Ions are released along a line (please note here we do not use Heed). The drift lines
of ions are shown. For a single-drift mesh setup, ions from Region 2, which represents the gap between the drift mesh and the ceiling of the test box are also collected on drift mesh. This increases the drift current for ions.}
\label{IonSingle}
\end{figure}

Our simulation begins with the electrostatic field within the 
given device using single drift mesh. 
As seen from Fig.~\ref{Field-Single}, in case of single drift mesh, 
the axial field in Region 2 (in this case, it is the region 
between drift mesh and the window), is negative.
The ions, thus, created in this region will be collected 
by the drift mesh, as shown in Fig.~\ref{Drift-Single}.
Therefore, the use of single drift mesh does not provide the 
correct estimation of the backflow fraction. 
To get a correct estimation of the backflow fraction, the contribution 
of these ions should be eliminated from the ionic current of the drift mesh.

\begin{figure}[hbt]
\centering
\subfigure[]
{\label{Field-Double1}\includegraphics[scale=0.1]{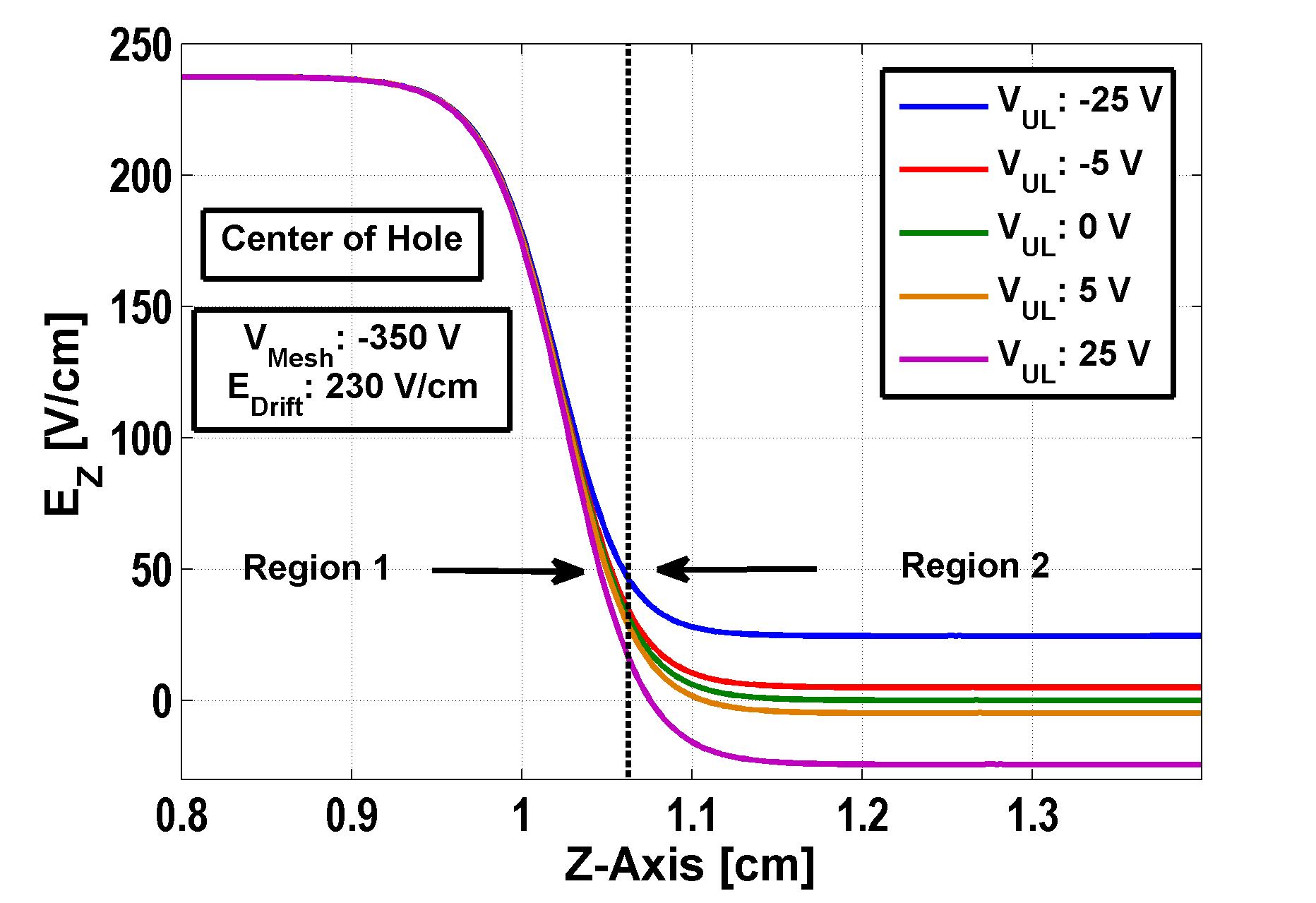}}
\subfigure[]
{\label{Field-Double4}\includegraphics[scale=0.1]{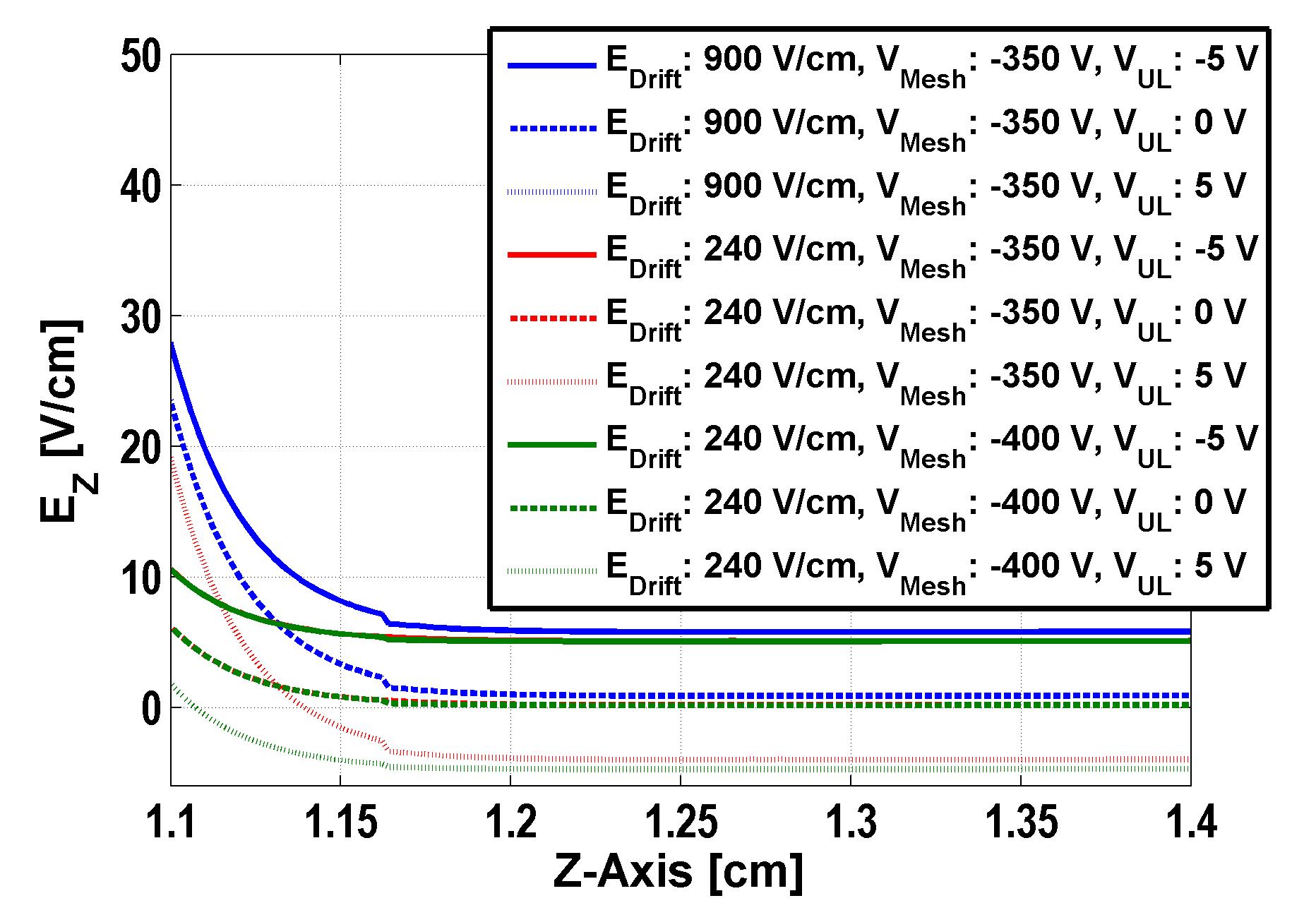}}
\caption{(a) Axial electric field in Region 2 and Region 3  
for a given set of $\mathrm{V_{UL}}= 0~\mathrm{V/cm}, \pm5~\mathrm{V/cm},
\pm25~\mathrm{V/cm} and +100~\mathrm{V/cm}$. The close-up of the 
region between $1.0~\mathrm{cm}~\mathrm{to}~1.2~\mathrm{cm}$ is shown in the 
inset. (b) For the same set of $\mathrm{V_{UL}}$, the axial electric field 
is plotted for different $\mathrm{E_{Amp}}$ and $\mathrm{E_{Drift}}$.}
\label{FieldDouble}
\end{figure}

In Fig.\ref{Field-Double1}, the axial electric field in Region 1 
and Region 2 is plotted, when there are two drift meshes.
The drift field in Region 1, is kept constant at 
$240~\mathrm{V/cm}$ whereas three 
different voltages have been applied to the $2^{nd}$ drift mesh ($V_U$) with respect 
to the $1^{st}$ drift mesh ($V_L$).
It can be inferred that the axial field in Region 1 ($\mathrm{E_{Drift}}$) remains
unaltered, as far as its direction is concerned.
$\mathrm{E_{R2}}$, on the other hand, has a change in direction due to the voltage
variation, which is expected.
It maintains a slightly positive value when there is no potential difference between
the two drift meshes.
This is probably due to the influence of the the field in Region 1.
In Fig.~\ref{Field-Double4}, we have presented the variation of $\mathrm{E_{R2}}$
due to a variation in amplification and drift fields.
It is apparent from the plot that the influence of the change in
amplification field is negligible and that of the drift field on $\mathrm{E_{R2}}$
is significant, but limited to a very small zone close to the boundary
of regions 1 and 2.

Therefore, only a positive $\mathrm{E_{R2}}$, i.e., a negative $\mathrm{V_{UL}}$,
may help in making the described laboratory device
representative of a TPC, as far as IBF is concerned.
In this context, the effect of $\mathrm{E_{R2}}$ on $\mathrm{I_C}$ and $\mathrm{I_M}$
needs to be mentioned here, since they will ultimately determine the estimate of IBF.

\textbf{Positive $\mathrm{E_{R2}}$}

Region 1: Ions produced likely to be collected by 1st drift mesh (desired);
Electrons takes part in avalanche, as usual (desired).
Region 2: Ions likely to be collected by 2nd drift mesh (desired);
Electrons likely to enter Region 1 and add to the count of electrons taking part
in avalanche.
This may lead to an increase in both $\mathrm{I_M}$ and $\mathrm{I_C}$, but
unlikely to misguide IBF measurement (not undesired).
So, both $\mathrm{I_C}$ and $\mathrm{I_M}$ are likely to increase for a positive
$\mathrm{E_{R2}}$.

\textbf{Negative $\mathrm{E_{R2}}$}

Region 1: Ions will be collected by Drift 1; Electrons take part in avalanche, as usual
(both desired).
Region 2: Ions drift towards Drift 1 leading to and increase in $\mathrm{I_C}$ and
erroneous estimate of IBF (undesirable);
Electrons gets collected by Drift 2 (desired).
So, $\mathrm{I_C}$ is likely to increase slightly, while $\mathrm{I_M}$ likely to decrease
for a negative $\mathrm{E_{R2}}$.

\begin{figure*}
\centering
{\label{IonEfficiency}\includegraphics[scale=0.2]{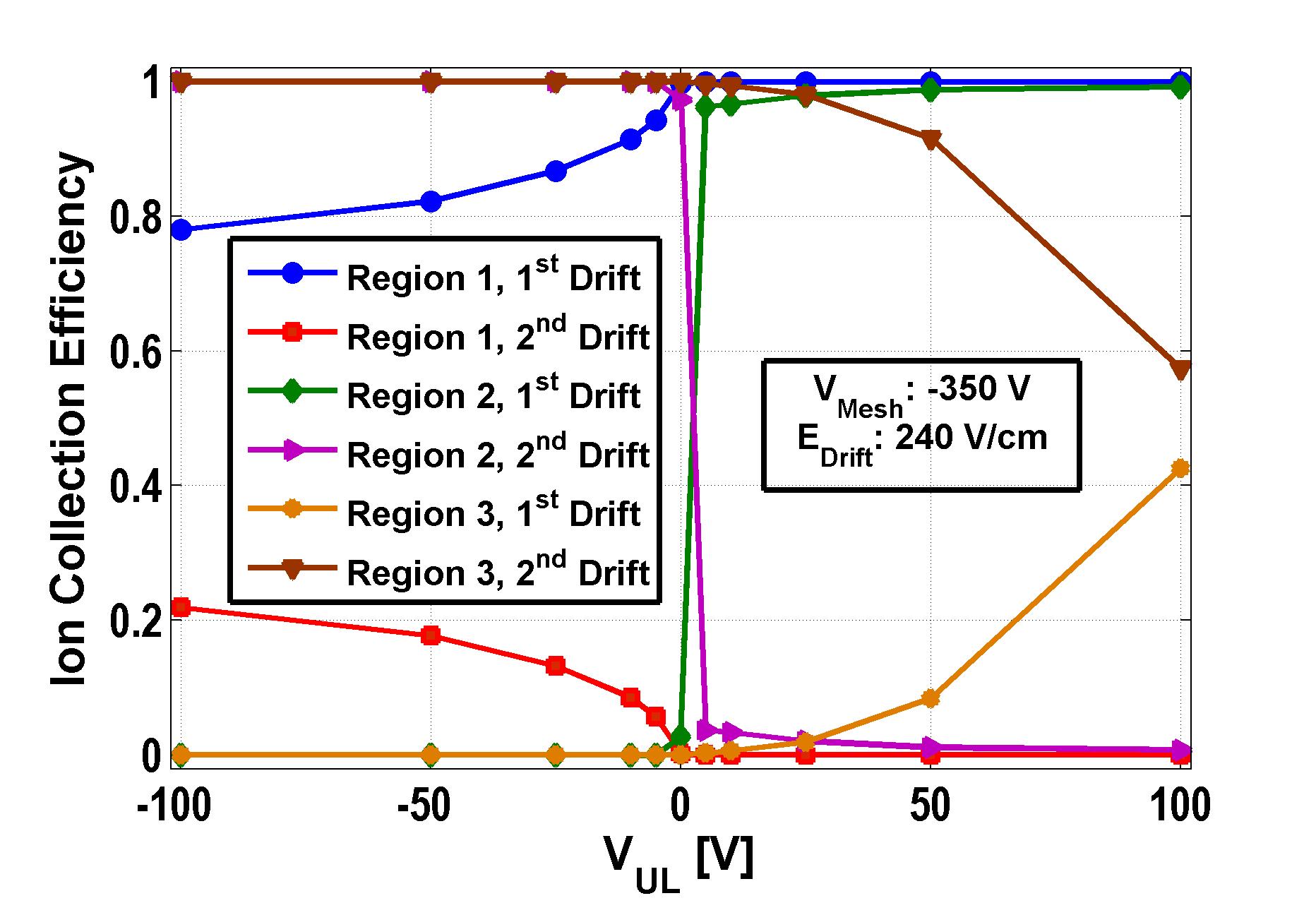}}
\caption{ Variation of ion collection efficiency of $1^{st}$ and $2^{nd}$ drift mesh 
with $\mathrm{V_{UL}}$.}
\label{DriftDouble}
\end{figure*}

To substantiate the preceding qualitative picture, the ion collection
efficiencies of the two drift meshes have been studied next,
as shown in Fig.9.
When $\mathrm{V_{UL}}$ is $0~\mathrm{V}$, 
the field ratio between Region 1 and Region 2 is tending to 
infinity and, thus, 
$100\%$ of ions from Region 1 are collected on $1^{st}$ drift mesh. 
And at the same time, $\sim3\%$ of ions from Region 2, which are created 
close to the $1^{st}$ drift mesh, are collected on that mesh. 
By making $\mathrm{V_{UL}}$ positive, the negative field 
in Region 2 pushes the ions towards the $1^{st}$ drift mesh. 
For example by making the $\mathrm{E_{UL}}$ $-5~\mathrm{V/cm}$, $\sim97\%$ 
of ions from Region 2 end their journey to the $1^{st}$ drift mesh.
On the other hand, if $\mathrm{E_{UL}}$ is positive, i.e., 
the $2^{nd}$ drift mesh is more negative than the $1^{st}$ one, 
the ions are attracted towards the $2^{nd}$ drift. 
But, in this case the field ratio between Region 1 and 2 is such that, 
some ions from Region 1 are also attracted by the $2^{nd}$ drift mesh. 
With the increase of positive field in Region 2, the number of ions
that goes from Region 1 to Region 2, increases. 
For example, when $\mathrm{E_{UL}}=5~\mathrm{V/cm}$, 
then, $\sim6\%$ of ions from Region 1 go towards Region 2 and are 
finally collected on the $2^{nd}$ drift mesh.
At the same time, $\sim100\%$ of ions in Region 2 also end their journey
on the $2^{nd}$ drift. 
If we increase $\mathrm{E_{UL}}$ to $10~\mathrm{V/cm}$, then $\sim9\%$ of 
ions from Region 1 end their journey on $2^{nd}$ drift mesh.
Therefore, though, a more negative voltage on the $2^{nd}$ drift mesh 
with respect to $1^{st}$ one is preferable, precise estimation of IBF is
poised on a delicate balance.

\subsection{Experimental Verification}

\begin{figure}[hbt]
\centering
\subfigure[]
{\label{ArIso-IC}\includegraphics[scale=0.3]{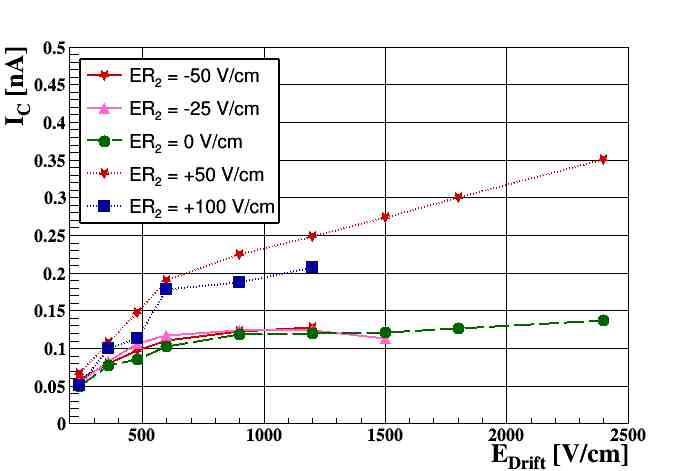}}
\subfigure[]
{\label{ArIso-IM}\includegraphics[scale=0.3]{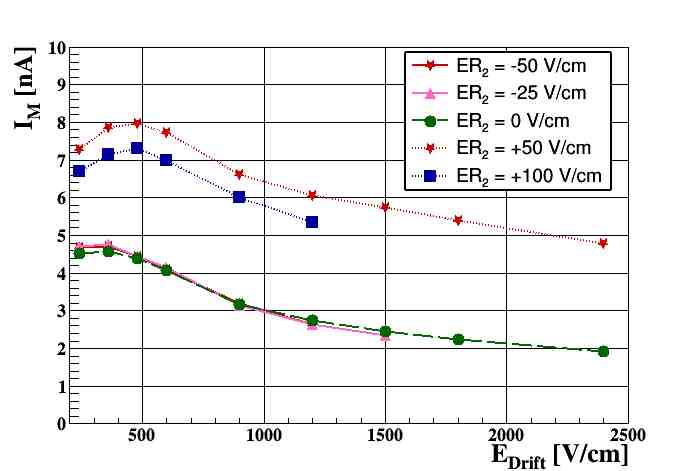}}
\subfigure[]
{\label{ArIso-IBF}\includegraphics[scale=0.3]{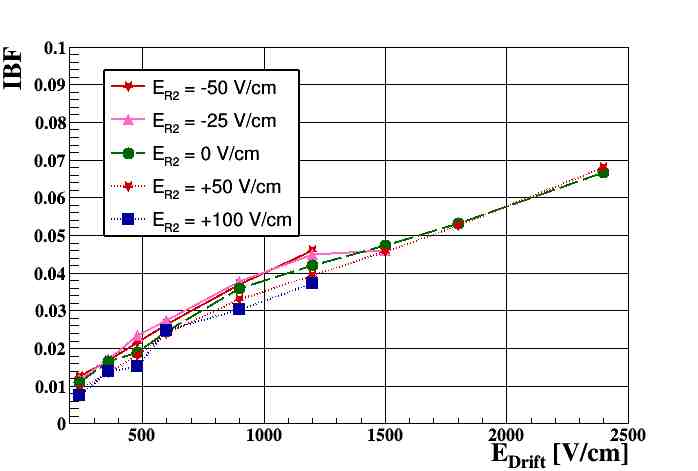}}
\caption{For Ar-Isobutane gas mixture (a) Variation of $\mathrm{I_C}$ with field ratio, (b)
Variation of  $\mathrm{I_M}$ with field ratio, (c) Variation of IBF with field ratio.}
\label{ArIso}
\end{figure}

\begin{figure}[hbt]
\centering
\subfigure[]
{\label{T2K-IC}\includegraphics[scale=0.3]{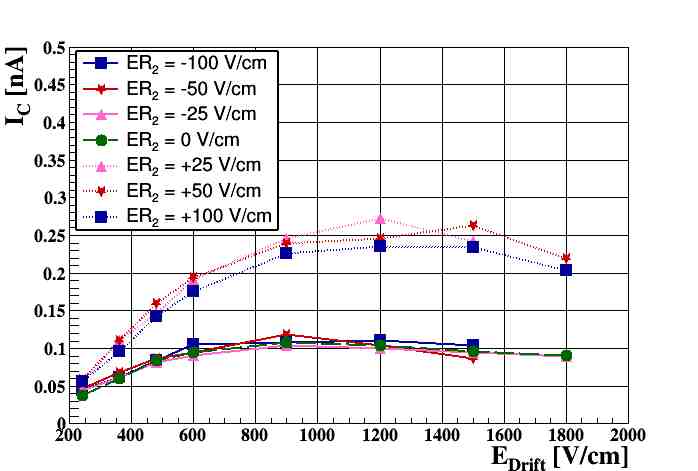}}
\subfigure[]
{\label{T2K-IM}\includegraphics[scale=0.3]{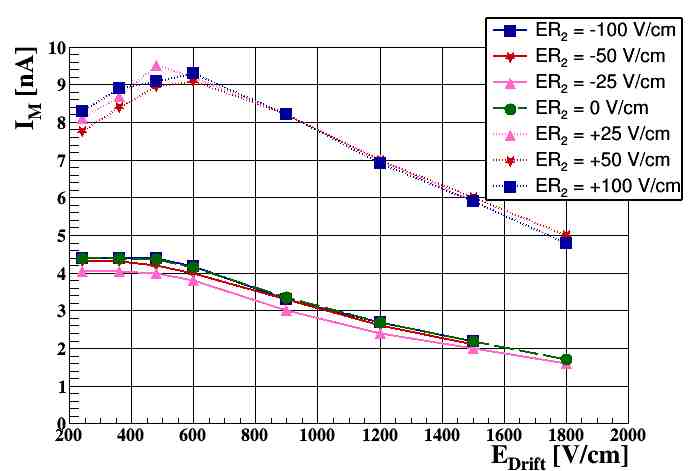}}
\subfigure[]
{\label{T2K-IBF}\includegraphics[scale=0.3]{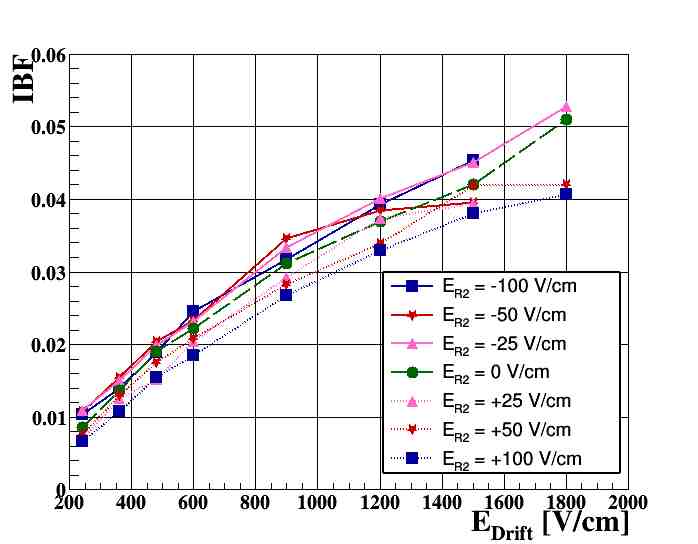}}
\caption{For T2K gas mixture (a) Variation of $\mathrm{I_C}$ with field ratio, (b)
Variation of  $\mathrm{I_M}$ with field ratio, (c) Variation of IBF with field ratio.}
\label{T2K}
\end{figure}

In order to verify the mentioned understanding of the processes, a series
of measurements of $\mathrm{I_C}$ and $\mathrm{I_M}$ were carried out for a bulk Micromegas
as described in section 3.1.
In order to understand the effect of gas composition, two popular gas
mixtures at room temperature and pressure were used for two separate
sets of measurements.
For these measurements, $\mathrm{E_{Drift}}$ has been varied, while, 
$\mathrm{E_{Amp}}$ remains constant.
The IBF for these gas mixtures were estimated from the measured
values of $\mathrm{I_C}$ and $\mathrm{I_M}$.
The results are presented in Figs. \ref{ArIso} and \ref{T2K}.
Figure \ref{ArIso} (a), (b), (c) respectively represents variation of
$\mathrm{I_C}$, $\mathrm{I_M}$ and IBF with field-ratio, as obtained for
Ar+isobutane gas mixture.
Similarly, Fig. \ref{T2K} represents those for Ar+isobutane+CF4.

If Figs. \ref{ArIso-IC} and \ref{T2K-IC} are considered, it is observed that
$\mathrm{I_C}$ increases slightly (5\%) when $\mathrm{E_{R2}}$ is negative w.r.t. its value
when $\mathrm{E_{R2}}$ = 0.
The increase in $\mathrm{I_C}$ is significant (50\%) when $\mathrm{E_{R2}}$ is positive,
especially for small field ratios, i.e., large drift fields.
Comparing Figs. \ref{ArIso-IM} and \ref{T2K-IM} we can conclude that
$\mathrm{I_M}$, for both the gas compositions, reduces slightly (less than 10\%),
or maintains its value when $\mathrm{E_{R2}}$ is negative, in comparison to the
$\mathrm{E_{R2}}$=0 case.
$\mathrm{I_M}$ increases significantly (even up to 100 \%) when $\mathrm{E_{R2}}$ is positive.
It should be mentioned here that the experimental observation matches
the understanding we developed from numerical simulations presented
in section 4.1.
The variations of the resulting IBF for these two gases have been
presented in Figs. \ref{ArIso-IBF} and \ref{T2K-IBF}.
Here, the results for Ar+isobutane and T2K gases are consistent with the
presented interpretation.
Thus, for these two gas mixtures, as shown in Figs. \ref{ArIso-IBF} and \ref{T2K-IBF},
IBF for $\mathrm{E_{R2}}$=0 is more than those for $\mathrm{E_{R2}}$ positive, and less
than those for $\mathrm{E_{R2}}$ negative.

\begin{figure*}
\centering
\subfigure[]
{\label{IBF}\includegraphics[scale=0.4]{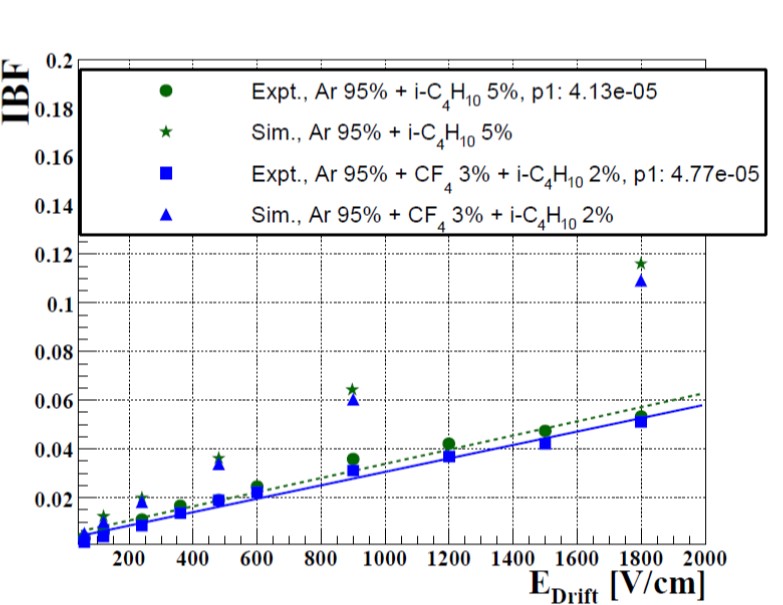}}
\subfigure[]
{\label{ArIsovsT2K}\includegraphics[scale=0.4]{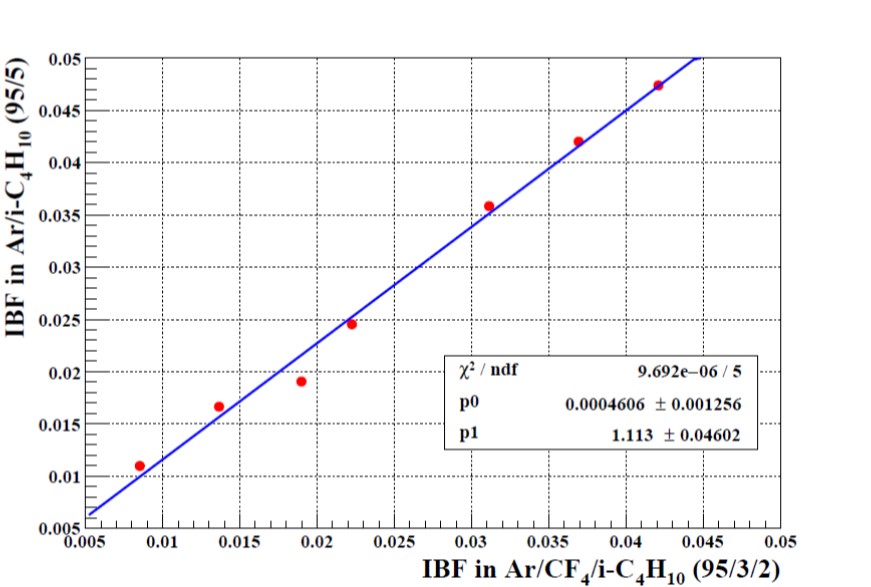}}
\caption{Variation of (a) IBF with field ratio for different gas compositions, (b) IBF
of T2K with respect to Ar-Isobutane.}
\label{Ratio}
\end{figure*}

Finally, in Fig. \ref{Ratio}, the variations of IBFs in the two gas compositions with drift field have been presented along with comparisons 
with numerical simulations (for $\mathrm{E_{R2}}$=0).
In Fig. \ref{IBF}, the IBFs for the gas compositions have been compared with their corresponding simulated estimates.
It is clear that the simulated values are almost twice the experimentally
measured values.
We believe that this mismatch is due to the fact that the geometry of numerical model is different from that of the actual detector.
For the numerical model, we have used the nominal design values, while deviations are known to occur during fabrication of the actual device.
We will look into this possibility in near future.
In Fig \ref{ArIsovsT2K}, the IBFs of Ar+Isobutane gas have been compared w.r.t the IBF of the T2K gas.
The gradients of the resulting lines are found to be 1.113.
Using Magboltz, it can be shown that the ratio of coefficients of transverse diffusion for
Ar+Isobutane to T2K gas mixtures is 0.9535.
Since, for a given field ratio, the IBF is expected to vary with inverse of the square of the coefficient of
transverse diffusion of a gas, the expected gradient is 1.094.
It may be mentioned here that, while estimating the transverse diffusion coefficient, the amplification field has been considered to be 27 kV/cm for both the cases.

\section{Conclusion}

The aim of this work has been to optimize an experimental setup to measure the ion backflow fraction (IBF) in a test-box so that it may closely represent the IBF measurement in a Time Projection Chamber. 
The conventional setup for measuring the backflow fraction involve inaccuracy due to the presence of the additional 
ions in the region between the drift mesh (cathode) and the test box window. 
The present work involves the measurements of the cathode and the mesh currents
for two Ar-based gas compositions and estimate corresponding ion backflow fraction 
of a bulk Micromegas detector using experimental setup comprising 
two drift meshes.
The Garfield simulation framework has been employed to find a guide line for
creating the field
configuration
so that an equivalent environment of the TPC can be established using 
the present double drift-mesh experimental setup.
The understanding developed from the numerical estimations and the measurements
agree reasonably well.
However, for IBF, numerical estimates have turned out to be larger than the measured values.
It is probably due to the fact that the geometry of the actual detector is different from the one modelled numerically, which followed the nominal specifications.

\section{Acknowledgment}
We thank Pradipta Kumar Das and Saibal Saha for their technical assistance during the experiment. 
This work has partly been performed in the framework of the RD51 
Collaboration.
We happily acknowledge the help and suggestions of the members of the 
RD51 Collaboration.
We thank our colleagues from the LCTPC collaboration for their help 
and suggestions.
Finally, we are thankful to our respective Institutions for providing 
us with the necessary facilities and IFCPAR/ CEFIPRA (Project No. 4304-1) 
for partial financial support.

\end{document}